\documentclass[12pt]{article}
\usepackage{epsfig}
\usepackage{graphicx}
\begin{document}
\vspace{.5in}

\begin{center}
{\large {\bf Computation of Horizontal Correlation of Sound in Presence of Internal Waves in Deep Water and Long Distances}}

\vspace{.3in}
\end{center}

\begin{center}
\vspace{.15in}

John L. Spiesberger\\ 
({\it Department of Earth and Environmental Science, University of Pennsylvania, Philadelphia, United 
States})\\
johnsr@sas.upenn.edu\\
\vspace{.15in}

\end{center}

\begin{center}
ABSTRACT
\end{center}
\vspace{.2in}

Numerical solutions are given for a parabolic approximation of the
acoustic wave equation at 200 and 250 Hz in two and three spatial dimensions
to determine if azimuthal coupling in the cross-range coordinate significantly affects
horizontal correlation in the presence of internal gravity waves in the sea. No evidence for
coupling is found for 
distances of 4000 km and less.  This implies
that accurate solutions are possible using computations
from uncoupled vertical slices. Shapes of horizontal correlation
are closer to shapes given by two theories than at lower frequencies.
\begin{flushleft}

\end{flushleft}

\begin{flushleft}
{\bf I. INTRODUCTION}
\end{flushleft} 

The horizontal correlation of low frequency sound in the deep ocean is important for practical problems including
how long an array can be built to yield gain in signal-to-noise ratio via beamforming. Applications include Navy detection and location
systems and international programs for the detection and location of nuclear blasts in the sea.  It has long been believed that
internal gravity waves may set the limit for horizontal correlation in these circumstances, and comparison between models and
data seem to back this up \cite{npal1,voron1,andrew,vera,spies_hoke}. The models in these studies did not solve a three spatially dependent
wave equation (3D) because of the very large computer demands in doing so.  Instead, the models computed
two-dimensional solutions of a wave equation along vertical slices through a three dimensional sound speed field.  This is known as a
N x 2D approach and is much more efficient than computing a 3D solution. In any case, the fact that the N x 2D solutions resemble the measurements
indicates a 3D solution may not be needed in such cases. The purpose of this study is to see if there is any significant difference between
a N x 2D solution and a 3D solution at frequencies of 200 and 250 Hz and distances up to 4000 km. Previous numerical studies found no significant
difference up to 150 Hz \cite{spies2007, spies2010}, which is consistent with analysis of signals near 75 Hz and a few thousand kilometers distance
\cite{npal1,voron1,andrew,vera}.   Comparisons between 2D and 3D solutions have not been made at higher frequencies such as
250 Hz where experimental measurements of horizontal correlation have been made \cite{spies_hoke}.
Due to the unusual availability of a large amount of computer time at a supercomputer center, the comparison at these higher
frequencies was made.  There are dozens of experiments up to 250 Hz and at basin-scales 
where horizontal arrays have collected data, so the current study is relevant for considering whether or not a 3D solution is needed to predict
horizontal correlation.

Solution of the linear acoustic wave equation is impractical on a supercomputer at 250 Hz and 4000 km distance. Instead, a parabolic 
equation (PE) is solved that is barely practical to implement.  There are many PEs that could be used for this calculation, and the one used
here is very accurate in the vertical dimension, yielding accurate travel times of all acoustic paths at thousands of kilometers distance,
regardless of the launch angle of the paths at the source \cite{c0}. At long distances, an accurate PE is needed in this application because
these launch angles go from roughly -15 to +15 degrees.  On the other hand, the horizontal correlation of sound is of order 1 km
at a few hundred Hertz
and order 1000 km distance, so the effect is associated with  a horizontal angle of about $1/1000 = 0.001$ radians.
All of the PE approximations are reported to be valid at small angles, and the one we choose is a standard small angle approximation for the
horizontal component  \cite{mpl}. No PE approximations would get the phase difference right at hydrophones separated by fractions of kilometers or more
at distances of several thousands of kilometers, not even the sound-speed insensitive approximation if it could
be made to work in both the horizontal and vertical planes with coupling. But it does not seem important to 
get the absolute phase differences right to get an accurate estimate of the horizontal correlation due to 
{\it fluctuations} of the sound speed field. Its the {\it relative changes in the phases} with time at two points 
that matter for computing
horizontal correlation.  It is not known through any mathematical theorem if any existing PE approximation is accurate enough 
to yield reliable estimates of the sought-after
effects. Another possible approach is to utilize the Thomson-Chapman wide angle PE approximation in the
vertical and horizontal planes \cite{tc}. It is not known if it is any more accurate than the approximation used here.
With the wide-angle approximation, for example, 
travel times of some multipath at 1000 km can be in error by order one second because the single reference speed
needed for the approximation makes the travel time right for only one equivalent vertical launch angle. 
In summary, the approximations used here are reported to be valid for the magnitude of the effects being investigated. 
Results from this paper could be checked in the future if someone invents a 
computationally practical method 
yielding more accurate solutions for the wave equation.

\begin{flushleft}
{\bf II. LIMITED HORIZONTAL DOMAIN AND CONVERGENCE}
\end{flushleft}

The calculations follow
the same methods used before \cite{spies2007,spies2010}, including the method by 
Smith \cite{smith} that was followed meticulously in all
respects to make the problem computationally practical by confining the numerical solution to a limited
horizontal domain (Fig. 1). 
We elaborate on two additional details not previously reported.
Firstly, Smith explains the computational domain near the source must first be computed over
the entire 360 degrees of azimuth for a few acoustic wavelengths. Afterward, the computational
domain may be restricted horizontally (e.g. Fig. 1 in Ref. \cite{spies2010}). We verified that the size of the 360 degree domain
was large enough so that further enlarging had insignificant affect on the computed solution at the receivers.
Secondly, we validated numerically that the rather narrow-looking damping regions on the periphery were wide enough
to attenuate significant reflections from the horizontal boundaries of the domain (e.g. Fig. 1 in Ref. \cite{spies2010}).  
The overall horizontal extent of the domain was set to four Fresnel radii as that was wide enough to not
contaminate with boundary reflections the central computational region for assessing horizontal correlation 
(Eq. 10 in Ref. \cite{spies2010}).  Numerical convergence of the PE at 200 and 250 Hz was obtained 
with horizontal and vertical grids
of 50 m and 1 m respectively (Table I). The PEs that were solved are exactly the same as shown in \cite{spies2007,spies2010}
except a typographical error is corrected here by  placing a plus sign after the $n^2$ term in Eq. 3.

\begin{figure}[ht]
\begin{center}
\psfig{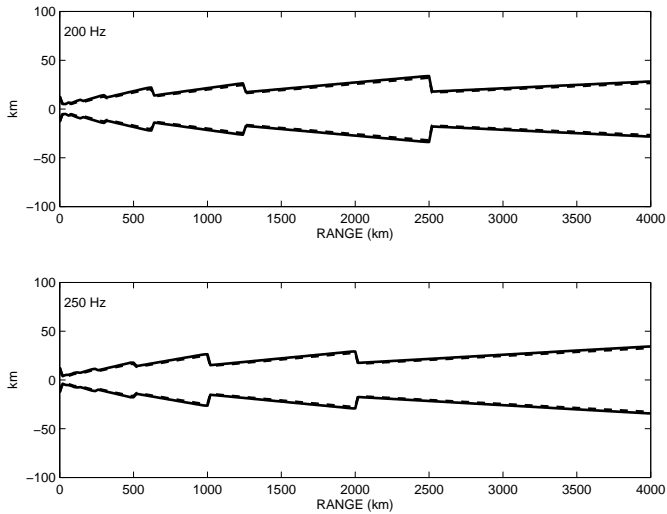}
\vspace{.1in}
\caption{Horizontal domains for computing parabolic approximation
of the acoustic wave equation at 200 and 250 Hz.  Horizontal and
vertical axes indicate distances along and perpendicular
to  the central geodesic.
The central geodesic is that between the source and receiver
on the midpoint of the horizontal array at 4000.74 km.  Values of the
acoustic field are undamped within dashed lines.  Damping occurs
between dashed and solid lines.}
\end{center}
\label{f:cxy_domain_200_250Hz}
\end{figure}

\begin{table}[th]
\begin{center}
\caption{Range step and number of depths needed in computational
grid for 3D parabolic approximation to obtain convergence of solution at indicated
frequencies.  The number of depths are those in the computational domain,
which extends from 0 to 8000 m depth.  Approximate times for a single uncoupled and coupled azimuth run are given
for an IBM Power5+ processor at 1.9 GHz. Computations at 200 and 250 Hz require about
2.5 GB of memory per cpu.  Codes are written
in Fortran and make use of IBM's Engineering Scientific Subroutine 
Library (ESSL) and Mathematical Acceleration Subsystem (MASS) 
libraries. Jobs were run using the Message Passing Interface (MPI) with 16 cpus.} 
{\begin{tabular}{r|rrrr} 
\multicolumn{1}{c|}{f(Hz)}&
 \multicolumn{1}{c}{$\Delta r$ (km)}&
   \multicolumn{1}{c}{\#}&
    \multicolumn{2}{c}{Computer CPU Hours}\\
\multicolumn{1}{c|}{}&
  \multicolumn{1}{c}{}&
   \multicolumn{1}{c}{Depths}&
 \multicolumn{1}{c}{Uncoupled}&
   \multicolumn{1}{c}{Coupled}\\ \hline
200&0.05&$2^{13}$&1200&2000\\
250&0.05&$2^{13}$&1600&2400\\ 
\end{tabular}}
\end{center}
\label{t:cputime}
\end{table}

\begin{flushleft}
{\bf III. RESULTS}
\end{flushleft}

At 200 and 250 Hz, and at all distances between 1000 and 4000 km, there are no significant differences between 2D and 3D solutions
of the PEs (e.g. Fig. 2). Computations at 200 and 250 Hz required about 200,000 cpu hours
on an IBM Power 5+ processor at 1.9 GHz. Statistical convergence was obtained with 17 snapshots 
of the internal wave field
at 200 Hz, and 37 snapshots at 250 Hz. For each snapshot, the acoustic field was computed on an array
of length 14 km. 
The interval between snapshots was one-half day where the 3D internal wave field
was temporally evolved using the linear dispersion relation. The results in  
Fig. 2 have about 1727 degrees of 
freedom (37 snapshots $\times$ 14 km array length/0.3 km correlation length). 

\begin{figure}[ht]
\begin{center}
\psfig{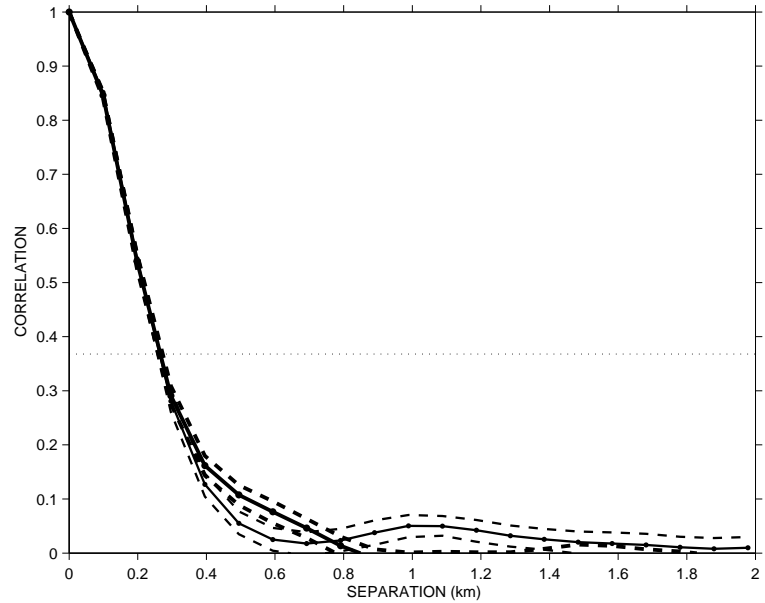}
\vspace{.1in}
\caption{95\% confidence limits via the bootstrap \cite{efron}
for horizontal correlation  at 250 Hz at 500 m depth and 4000 km
distance from source for uncoupled azimuth (dark lines) and coupled azimuth (light lines)
solutions of the 3D wave equation.
The dotted
line is at $e^{-1}$.}
\end{center}
\label{f:coupled_uncoupled_250Hz}
\end{figure}

Voronovich and Ostashev\cite{voron3,vo2} developed theories to estimate when effects
of azimuthal coupling were significant.  Their theories have not been used
to draw boundaries in frequency/transmission-distance space that designate when the 3D solution
is required.  The second of their theories \cite{vo2} appears to be the most advanced,
and they report a distance of validity up to about 1000 km. The numerical results here might 
be used to compare with that or other theories.

The numerical calculations for the shape of horizontal correlation are sometimes parametrized by the
form $\exp(-y/y_*)^p$ where $y$ is horizontal separation and $y_*$ and $p$ are constants derived from
theory or fit to the numerical calculations. The first theory \cite{stoughton} yields $p=1.5$ and the 
second \cite{vo2,voron2}
yields $p=2$.  The numerical calculations at 200 (not shown) and 250 Hz (Fig. 3) yield values of $p$ from about
1.4 at 500 km to about 1.9 at 4000 km, so they do not help confirm one theory over another. 
Numerical calculations at 150 Hz and lower \cite{spies2007,spies2010} yield values of $p$ between about 1 and 1.4,
which are inconsistent with both theories. It is not known why
the numerical values come closer to theoretical results at higher frequencies.

\begin{figure}[ht]
\begin{center}
\psfig{figure=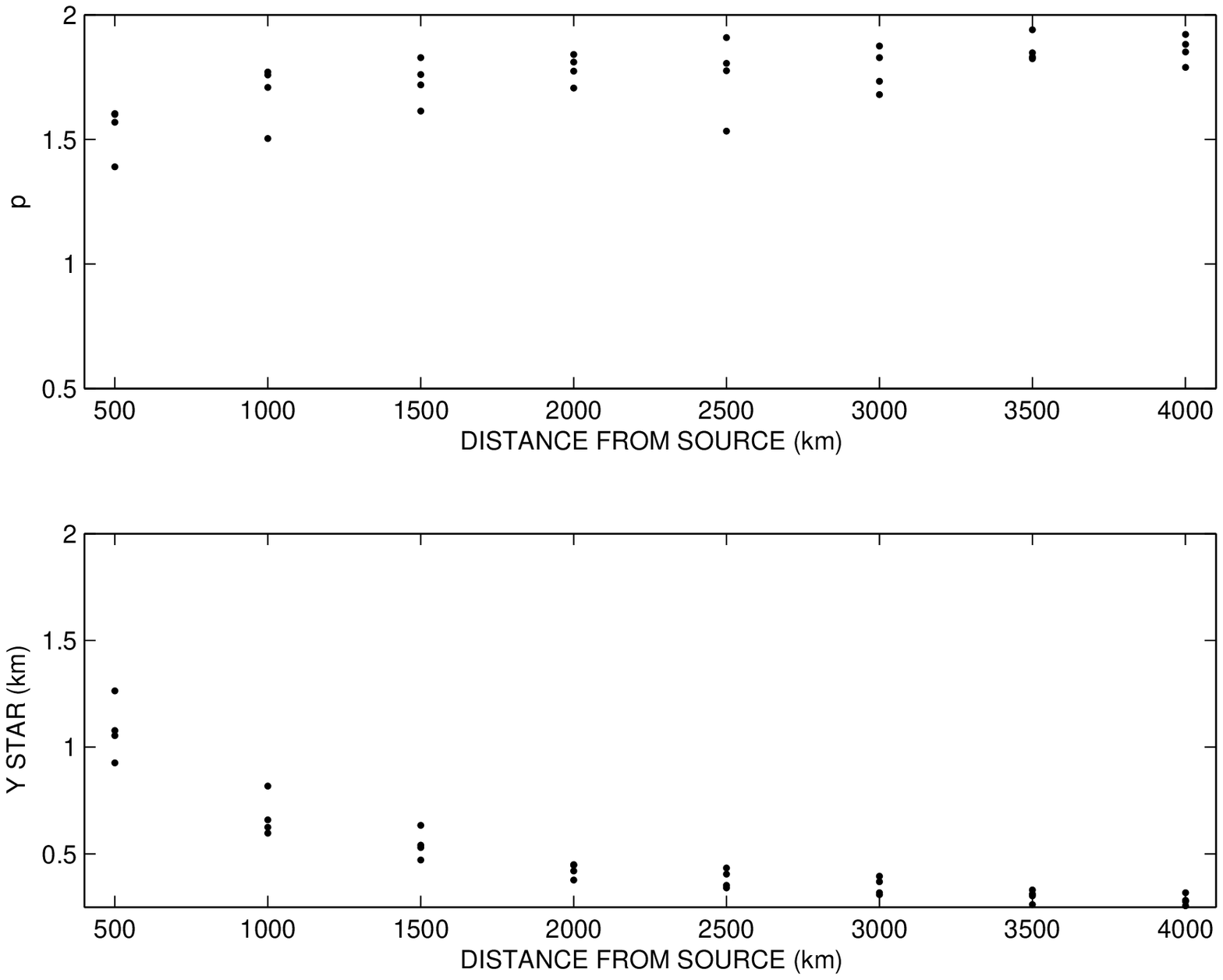,width=4.in}
\vspace{.1in}
\caption{Least-squares fit of numerically-computed horizontal correlation at 250 Hz with $\exp[-(y/y_*)^p]$
where minimum is found over all values of $p$ and $y_*$.  Results are given for indicated distance
from source at each of four depths (100, 500, 1000, and 3000 m). The fit is made only to model values
with correlation exceeding 0.3.}
\end{center}
\label{f:fitshape250}
\end{figure}

\begin{flushleft}
{\bf ACKNOWLEDGMENTS}
\end{flushleft}

This research was supported by the Office of Naval Research contracts
N00014-06-C-0031,  N00014-10-C-0480, N00014-12-C-0230, and by a grant of computer time from the
DOD High Performance Computing Modernization Program at the Naval Oceanographic Office.
I thank Drs. Michael Brown and Kevin Smith for many suggestions.

\newpage 
\begin{flushleft}

\end{flushleft}

\end{document}